\documentclass[11pt]{article}

\usepackage{jcappub}
\usepackage{graphicx,color,soul}
\usepackage{latexsym}
\usepackage{amsmath,amssymb}        % amssymb includes amsfonts

\title{PBH mass growth through radial accretion during the radiation dominated era}

\author[1,2]{F. D. Lora-Clavijo}
\affiliation[1]{Instituto de Astronom\'ia, Universidad Nacional Aut\'onoma de M\'exico, 
	AP 70-264, Distrito Federal 04510, M\'exico.}

\author[2,3]{F. S. Guzm\'an,}
\author[2]{A. Cruz-Osorio.}

\affiliation[2]{Instituto de F\'{\i}sica y Matem\'{a}ticas, Universidad
              Michoacana de San Nicol\'as de Hidalgo. Edificio C-3, Cd.
              Universitaria, 58040 Morelia, Michoac\'{a}n,
              M\'{e}xico.}

\affiliation[3]{Department of Physics and Astronomy, 
	University of British Columbia.
	6224 Agricultural Road, Vancouver BC, Canada, V6T1Z1.}

\emailAdd{fdlora@astro.unam.mx}
\emailAdd{guzman@ifm.umich.mx}
\emailAdd{alejandro@ifm.umich.mx}

\date{\today}

\abstract{
We model the radial accretion of radiation on Primordial Black Holes (PBH) by numerically solving Einstein's equations coupled to an ultrarelativistic ideal gas with equation of state $p=\rho/3$. We calculate the final mass of a black hole by the integration of the accreted radiation energy density  during the leptonic era between $t\sim10^{-4}s$ to $t\sim 10^2s$ after the Big Bang. Our results indicate that small PBHs with initial masses between $10^{-4}$ to $1M_{\odot}$ may grow up to hundreds of solar masses, and thus can be SMBH seeds. On the other hand, PBHs formed at $t\sim 1s$ with initial mass between 900 and $\sim 980M_{\odot}$, by the time $t\sim 100s$ show masses of $10^4$ to $10^6M_{\odot}$ which are masses of seeds or already formed SMBHs. The fact that we consider only radial flow implies that our results work well as limiting cases, and it is expected that under more general scenarios the accretion rates may change significantly.  Nevertheless we show that it is possible that SMBHs can be PBHs that grew due to the accretion of radiation.}


\begin{document}

\maketitle

% ----->     INTRODUCTION     <-----

\section{Introduction}

The problem of formation and evolution of supermassive black holes (SMBHs) in the center of a considerable amount of elliptic and disk galaxies remains unsolved. The black hole growth is usually related to its coexistence with the surrounding matter, both baryonic and dark matter. Some models consider these black holes are the result of the evolution of seed black holes (\cite{Eisenstein1995,Bullock2004,Heger2003,GuzmanLoraSph1,GuzmanLoraSph2,CPepe,Daniel}) of various initial masses that grow through accretion.

Motivated by the question of how could SMBH seeds have formed, it could be that these black holes are primordial black holes (PBHs) that grew through accretion of radiation until they reached the masses corresponding to SMBH seeds. PBHs are assumed to be black holes formed in the cosmological context via the gravitational collapse of overdense regions of primordial density fluctuations during the early universe \cite{Zeldovich,Hawking}. PBHs are important because they may have important observational implications, for instance: small black holes could have evaporated via Hawking radiation adding to the $\gamma$ ray background very short gamma ray bursts \cite{Cline}, 
binary systems of primordial black holes could produce gravitational radiation \cite{Nakamura}, 
if there were a large number of PBHs they would contribute significantly to the cosmic density parameter $\Omega$, among others.

An important issue concerning PBHs is the formation mechanism. The formation is usually assumed to happen during a radiation-type dominated era (RDE) where the dominant equation of state corresponds to ultrarelativistic material. Some of the most discussed formation mechanims are for instance that PBHs formed at the QCD phase transition, where modest overdensity regions could have collapsed to form black holes \cite{qcd}, the collapse of cosmic string loops \cite{strings}, 
bubble collisions during the spontaneous symmetry breaking \cite{bubbles}, the collapse of domain walls \cite{Berezin}, and the collapse of matter during a stage where the pressure may have decreased implying a soft equation of state \cite{Khlopov}. The process of formation has been also extensively analyzed numerically, starting from an initial fluctuation that collapses to form the hole, which has provided important properties of both, the collapse itself and the mass, time-scales and density contrast required for the collapse \cite{Simulations}.

Even though the process of collapse is very interesting itself, studying the consequences of accretion of matter by PBHs is also important and specially relevant for the SMBH seed formation. Different models of PBH growth depend on the type of matter accreted and the hypotheses of each growth model. Early models are based on Bondi type of accretion that in first approximation do not consider the cosmic expansion, and study the accretion of radiation by PBHs \cite{Zeldovich}. 
Corrections to such model include the cosmic expansion and the construction of self-similar solutions for the accretion of barotropic fluids with equation of state $p=(\Gamma-1)\rho$ and $\Gamma=4/3$, and showed that the mass of a PBH cannot grow as fast as the universe \cite{Carr}, and that the growth of PBHs due to the accretion of radiation during the RDE is not significant. Further refined models include the analysis of solutions for the accretion of gas with various values of $1< \Gamma <2$, and it is shown  that accretion is in any case very small \cite{HaradaCarrA}.

The analyses of PBH accretion has also expanded to the accretion of dark energy \cite{Babichev}, and the possibility that SMBHs are the result of the accretion of quintessence fields by PBHs and found that such scenario is consistent with bounds on SMBH masses \cite{Bean}, although the analysis there does not consider the full evolution of the scalar field profile that could bring to runaway instability \cite{CMC} or prevent the scalar field to be accreted only partially in terms of the size of the wave packet \cite{UrenaLiddle,LoraGuzmanSF};  in \cite{Harada} also the accretion of a massless scalar field is shown to contribute with at most a factor of two to the PBH mass based on the non-linear solution of Einstein equations; in \cite{Mack}, the contribution of the accretion of dark matter to PBHs mass is analyzed and also found that PBHs may grow two orders of magnitude during the RDE. A recent and complete review of the PBH growth including various types of matter in found in \cite{CarrReview}.

Despite of the interesting possible scenarios of accretion of dark components and other matter fields by PBHs, in this paper we are particularly interested in the calculation of the accretion of radiation on PBHs by solving the Einstein-Euler system of equations numerically and measure the growth of the black hole's apparent horizon.  

In order to model the PBH+radiation system at a local scale, we consider a  Schwarzschild type of black hole described using horizon penetrating coordinates, immersed in a sea of radiation falling in radially. In order to study this system numerically we use a finite domain, imposing an artificial boundary far from the event horizon of the black hole, at a finite distance, where the inward flow of radiation is allowed. 
A condition we impose in our analysis, is that the cosmological particle horizon has to be much bigger than the black hole event horizon radius. This is an important condition that allows us to detach the cosmic expansion from the local accretion process at local scale \cite{Zeldovich}.

We solve numerically the Einstein-Euler system of equations for a fluid obeying an ideal gas equation of state in the limit of ultrarelativistic material, that is, assuming the rest mass density of the gas is much smaller than the total energy density of the gas and also assuming a radiation type of equation of state, which is the one used to model  the radiation during the RDE. We inject the radiation through the exterior boundary of the domain. The density of such ingoing gas corresponds to the mean density of the universe at a given time. We consider that the density of the universe goes as $\rho \sim 1/t^2$ during the RDE and in particular we study the time window $t \in [10^{-4}s,100s]$ within the leptonic era.

An important obstacle at this point is that the time and spatial scales change a number of orders of magnitude during such time window, which numerically becomes a significant problem, for instance if the black hole mass grows also orders of magnitude, so does its horizon and therefore the numerical domain; this is a reason why studies involving the formation of PBHs due to collapse of fluctuations cannot be carried out during arbitrarily large evolution time scales once the hole has formed. A strategy to study the PBH growth is to solve the coupled Einstein-Euler system for a radiation fluid, during a given -numerically tractable- lapse of time $t\in[t_a,t_b]$ using a prescribed mean density of the universe, during a time scale in which the mean density of the universe does not change significantly.

The result of solving the Einstein-Euler system assuming a constant in time asymptotic/environment density, is that the BH horizon mass grows linearly in time, that is, the accretion mass rate is constant; we take advantage of this result to study the growth of PBHs. The incorporation of the expansion of the universe is as follows. In order to track the evolution of space-time plus radiation system during the whole time window $t\in[t_0,t_f]$ (with e.g. $t_0=10^{-4}s$ and $t_f=100s$) considering a given environment density, we partitioned such time domain in a number of time intervals $t_0 < t_1 < t_2 < ... < t_{N-1} < t_f$, in each of which we assumed to hold the constant in time accretion rate found for the solution of the full Einstein-Euler system. Thus, within each interval $t\in[t_i,t_{i+1}]$ we assumed an initial BH horizon mass $M_i$, a mean density of the universe $\rho_i$ at time $t_i$, and estimated the final mass of the BH horizon at the end of the interval $M_{i+1}$ using the stationary growth of the horizon; then starting with the new value of the BH horizon mass $M_{i+1}$ we repeat the process until we arrive at $t_f$ where we find a final mass of the BH $M^{PBH}_{f}$. Notice that the cosmic expansion is considered  through the value of the density $\rho_i$. We make sure that we choose a sufficiently large value of $N$ such that the result becomes independent of the number of time intervals, that is, when the result is convergent up to machine precision. We call this procedure a sequence of stationary stages.

The paper is organized as follows. In section \ref{sec:equations} we write down the fully coupled system of equations describing the evolution of the ultrarelativistic gas and the black hole; in section \ref{sec:numerics} we describe the numerical methods we use to solve the system of equations. In \ref{sec:results} we present our results, in the first part the BH horizon growth in time where we show that it grows in a stationary way and in a second part we present a number of results related to the evolution of the PBH. Finally in \ref{sec:conclusions} we draw some conclusions.

% ----->     SECTION     <-----

\section{Evolution equations of the Einstein-Euler system}
\label{sec:equations}

In order to evolve the system of black hole plus the gas we consider  the 3+1 decomposition of the space-time for the evolution of the geometry, and a consistent description of the gas dynamics (see e.g. \cite{Alcubierre,Baumgarte}).

% ----->     Subsection: geometry     <-----

\subsection{Evolution of the space-time geometry}

In order to solve numerically the Einstein Field equations $G_{\mu \nu}=8\pi T_{\mu\nu}$, where $G_{\mu\nu}$ is the Einstein tensor and $T_{\mu \nu}$ is the energy momentum tensor, we use the 3+1 splitting approach of the general relativity and adopt the Arnowitt-Deser-Misner (ADM) formalism of the evolution equations. For a spherically symmetric space-time the line element can be written as follows

\noindent
\begin{eqnarray}
ds^2 = &-&(\alpha^2 - \gamma_{rr}\beta^2)dt^2 + 2\gamma_{rr}\beta dr dt + \gamma_{rr}dr^2 
\gamma_{\theta\theta} (d\theta^2 + \sin^2 \theta d\phi^2), \label{eq:lineelement}
\end{eqnarray}

\noindent where $\alpha$ is the lapse function, $\beta^i=(\beta,0,0)$  is the shift vector, $\gamma_{ij}=diag(\gamma_{rr},\gamma_{\theta\theta},\sin^2 \theta \gamma_{\theta\theta})$ are the components of the spatial 3-metric associated with the space-like hypersurfaces $\Sigma_t$ foliating the space-time and  $x^\mu=(t,r,\theta,\phi)$ are the coordinates of the space-time.

According to the 3+1 decomposition of the space-time, the components of the extrinsic curvature of slices $\Sigma_t$ the space-time is foliated with, are $K_{ij} = \frac{1}{2\alpha}[-\partial_t \gamma_{ij} + \nabla_i \beta_j + \nabla_j \beta_i]$, where $\nabla_i$ is the covariant derivative of the 3-dimensional spatial slices $\Sigma_t$. The non-trivial components of the extrinsic curvature consistent with (\ref{eq:lineelement}) are $K_{ij} = diag(K_{rr},K_{\theta\theta},\sin^2 \theta K_{\theta\theta})$. In general the ADM (Arnowitt-Deser-Misner) formulation of general relativity decomposes Einstein's equations into six evolution equations for $\gamma_{ij}$ and six more for $K_{ij}$, and additionally four constraints, the Hamiltonian constraint and three momentum constraints \cite{Alcubierre,Baumgarte}. In our case of spherical symmetry, written in spherical coordinate, the evolution equations are only four 

\begin{eqnarray}
          \partial_t \gamma_{rr} = &-&2\alpha K_{rr} + \beta^r \partial_r \gamma_{rr} + 2\gamma_{rr} \partial_r \beta^r, \nonumber \\ 
          \partial_t \gamma_{\theta \theta} = &-&2\alpha K_{\theta \theta} + \beta^r \partial_r \gamma_{\theta \theta}, \nonumber \\
          \partial_t K_{rr} = &-&\partial_{rr}\alpha + \frac{(\partial_r \gamma_{rr})( \partial_r \alpha)}{2\gamma_{rr}} + \frac{\alpha}{2}\left (\frac{\partial_r \gamma_{\theta \theta}}{\gamma_{\theta \theta}}\right )^2 
          - \alpha \frac{\partial_{rr} \gamma_{\theta \theta}}{\gamma_{\theta \theta}} + \alpha \frac{(\partial_r \gamma_{rr})(\partial_r \gamma_{\theta \theta})}{2\gamma_{rr} \gamma_{\theta \theta}} + 2\alpha\frac{K_{rr} K_{\theta \theta}}{\gamma_{\theta \theta}} 
           \nonumber \\&-& \alpha \frac{K_{rr}^2 }{\gamma_{rr}} + \beta^r\partial_r K_{rr} + 2K_{rr}\partial_r \beta^r 
           + 4\pi \alpha [(S-\rho_{ADM})\gamma_{rr} - 2S_{rr}], \nonumber \\  
           \partial_t K_{\theta \theta} = &-&\frac{(\partial_r \gamma_{\theta \theta})(\partial_r \alpha)}{2\gamma_{rr}} - \alpha \frac{\partial_{rr} \gamma_{\theta \theta}}{2\gamma_{rr}}  
           + \alpha \frac{(\partial_r \gamma_{rr})(\partial_r \gamma_{\theta \theta})}{4\gamma_{rr}^2} + \alpha \left [ 1 + \frac{K_{rr} k_{\theta \theta}}{\gamma_{rr}} \right ]  + \beta^r \partial_r K_{\theta \theta} \nonumber\\
           &+& 4\pi \alpha [(S-\rho_{ADM})\gamma_{\theta \theta} - 2S_{\theta \theta}].
           \label{eq:gammaevolve}
\end{eqnarray}  

\noindent and two constraint equations

\begin{eqnarray}
H&:=&{}^{(3)}R + K^2 - K_{ij}K^{ij} - 16\pi\rho_{ADM} = 0, \nonumber\\
M^r&:=&\nabla_j K^{rj} - \gamma^{rj} \nabla_j K - 8\pi j^r = 0, \label{eq:constrains} 
\end{eqnarray}

\noindent where $^{(3)}R$ is the scalar of curvature associated to $\gamma_{ij}$. Given $n^{\mu}$ is a 4-vector normal to the spatial hypersurfaces $\Sigma_t$ and $T_{\mu\nu}$ the stress energy tensor of  the matter field, in equations (\ref{eq:gammaevolve} - \ref{eq:constrains}), the quantities $\rho_{ADM}=n_{\mu}n_{\nu}T^{\mu\nu}$, $j^i = -\gamma^{ij}n^{\mu}T_{\mu j}$, $S_{ij}=\gamma_{i\mu}\gamma_{j\nu}T^{\mu\nu}$ and $S=\gamma^{ij}S_{ij}$ correspond to the local energy density, the momentum density, the spatial stress tensor and its trace respectively, measured by an Eulerian observer. These variables are obtained from the projection of the energy momentum tensor  $T_{\mu \nu}$ along the space-like hypersurfaces and the normal direction to such hypersurfaces. The gauge used during the evolution is such that we restore $\alpha$ and $\beta$ during the evolution in such a way that we keep $\gamma_{\theta\theta}$ constant in time and force the ingoing null rays at each point of the domain to satisfy $dt/dr = -1$ \cite{Thornburg}.

The key of the non-linear evolution is that aside of solving Einstein's equations it is required to solve the sources and the equations ruling the matter simultaneously. The evolution of the radiation model is ruled by the general relativistic Euler euqations described next.

% ----->     Subsection: Euler's equations

\subsection{Solution of Euler equations}

We model the radiation matter field as a perfect fluid with stress energy tensor $T^{\mu \nu} = (\rho+p)u^{\mu} u^{\nu} + pg^{\mu \nu}$, where $\rho$ is the energy density of the gas, $p$ its  pressure, $u^{\mu}$ is the 4-velocity of fluid elements and $g_{\mu \nu}$ are the components of the four-metric tensor. The complete set of Euler equations combines two conditions: 1) the conservation of the rest mass density $\rho_0$, that is, $\nabla_{\nu} (\rho_{0}u^{\nu}) = 0$, where the total energy density is $\rho=\rho_0(1+\epsilon)$, with $\epsilon$ is the internal energy of the gas, and 2) the Bianchi identity $\nabla_{\nu} (T^{\mu \nu}) = 0$, where $\nabla_{\mu}$ is the covariant derivative of the full 4-metric. 

The condition for the gas being ultrarelativistic consists in assuming that the rest mass density of the gas is much smaller than the total energy density $\rho_0 \ll \rho = \rho_0 (1+\epsilon)$. This condition implies that the conservation of rest mass energy density becomes an identity and Euler equations reduce only to the Bianchi identity (see e.g. \cite{Choptuik}). We model the radiation fluid with an ideal gas equation of state $p=(\Gamma-1)\rho$, where $\Gamma$ is the ratio between specific heats and takes the value $\Gamma=4/3$ for radiation.  

In order to write down the equations describing the evolution of the gas, in a way that standard numerical methods can be applied, we define conservative variables such that the set of ultrarelativistic Euler equations are written in a flux balance law form

\noindent
\begin{eqnarray}
\partial _{t} {\bf u} + \partial _{r} \big( {\bf F}^{r}  \big)   = {\bf S}, \label{eq:flux_conservative}
\end{eqnarray}

\noindent where ${\bf u}$ is a vector of conservative variables , ${\bf F}^{r}$ is the vector of fluxes and ${\bf S}$ are the sources. Explicitly these quantities read

\begin{eqnarray}
\nonumber {\bf u} &=& \left[ \begin{array}{l}
S_{r} \\
\tau   
\end{array} \right]=
 \left[ \begin{array}{l}
\sqrt{\gamma}(\rho + p) W^{2}v_{r} \\
\nonumber \\
\sqrt{\gamma}(\rho + p) W^{2}  - p
\end{array} \right], \\ 
\nonumber {\bf F}^{r} &=& 
\alpha \left[ \begin{array}{l}
 \Big(v^{r}-\frac{\beta^{r}}{\alpha}\Big)S_{r}  +  \sqrt{\gamma}p  \\
 \Big(v^{r}-\frac{\beta^{r}}{\alpha}\Big) \tau + \sqrt{\gamma}pv^{r} 
\end{array} \right], \\
\nonumber \\
{\bf S} &=& \
 \left[ \begin{array}{l}
 \alpha\sqrt{\gamma}T^{\mu \nu} g_{\nu \sigma } \Gamma ^{\sigma}_{\mu r} \\
\alpha\sqrt{\gamma}(T^{\mu t} \partial_{\mu}\alpha  - \alpha T^{\mu \nu} \Gamma^{t}_{\mu\nu}) 
\end{array} \right],\label{eq:sourcessphe}
\end{eqnarray}

\noindent for radial fluxes in spherical coordinates. In these expressions,  $\gamma=det(\gamma_{ij})$ is the determinant of the spatial metric, $\Gamma^{\sigma}{}_{\mu \nu}$ are the Christoffel symbols and  $v^{i}$ is the 3-velocity measured  by an Eulerian observer and defined in terms of the spatial part of the 4-velocity $u^{i}$ as $v^{i}=\frac{u^{i}}{W} + \frac{\beta^{i}}{\alpha}$, where $W$ is the Lorentz factor given by $W=\frac{1}{\sqrt{1-\gamma_{ij} v^{i}v^{j}}}$. Specifically, in the spherically symmetric case described with spherical coordinates, the non trivial component of the velocity is $v^i=(v^r,0,0)$, and thus $v^r=\frac{u^r}{W}+\frac{\beta^r}{\alpha}$, with $W = \frac{1}{\sqrt{1-\gamma_{rr}v^r v^r}}$.

In terms of the gas variables, the ADM matter sources  required for the evolution of the geometry  (\ref{eq:gammaevolve}) and the constraints (\ref{eq:constrains}) are as follows:
$\rho_{ADM} = (\rho +p)W^2 -p$, $j^r = (\rho +p) W^2 v^r$, $S_{rr} = (\rho+p) W^2 v_r v_r + \gamma_{rr}p$, $S_{\theta\theta} = \gamma_{\theta\theta}p$, $S = (\rho +p) W^2 v_r v^r + 3p$, which change in time and need to be constructed out of the conservative variables.

Finally, it is necessary to close the system of equations (\ref{eq:flux_conservative}-\ref{eq:sourcessphe}), using the equation of state $p=(\Gamma-1)\rho$.

% ----->     SECTION     <-----

\section{Numerical Methods for the evolution}
\label{sec:numerics}

\subsection{Evolution}

We solve the evolution equations for the geometry (\ref{eq:gammaevolve}) and matter (\ref{eq:flux_conservative},\ref{eq:sourcessphe}) on a discretized version of the spatial domain $r \in [r_{min},r_{max}]$. We use a uniformly discretized numerical grid and use the method of lines for the evolution of data from one time slice to the next one, with a third order Runge Kutta integrator \cite{Shu}. The right hand sides of the evolution equations for the geometry are discretized using fourth order finite difference stencils. On the other hand, Euler equations are discretized using a finite volume approach with a High Resolution Shock Capturing method,  that uses a minmod variable reconstructor, and the fluxes in ({\ref{eq:sourcessphe}) are calculated using the Harten, Lax, van Leer and Einfeldt (HLLE) approximate Riemann solver formula \cite{hlle}. The spectral structure of the Jacobian matrix of the gas system (\ref{eq:flux_conservative}-\ref{eq:sourcessphe}) consists of two eigenvalues 

\noindent
\begin{equation}
\nonumber \lambda_{\pm}=\alpha  \Big(v^{r}-\frac{\beta^{r}}{\alpha}\Big) + \alpha{\cal A} \pm \alpha \sqrt{ {\cal A} + (1-v^{2}) {\cal B}}, 
\end{equation}

\noindent where 

\noindent 
\begin{eqnarray}
\nonumber {\cal A}&=&\sqrt{\gamma}(1-v^{2})\frac{\partial p}{\partial S_{r}} /2, \\
\nonumber {\cal B}&=&\sqrt{\gamma}\Big(v^{r} \frac{\partial p}{\partial S_{r}}+\gamma^{rr}\frac{\partial p}{\partial \tau}\Big), 
\end{eqnarray}

\noindent which are the eigenvalues used into the HLLE flux formula. Since we evolve the conservative variables in (\ref{eq:flux_conservative}-\ref{eq:sourcessphe}) and the fluxes depend on both, the conservative variables ${\bf u}$ and the primitive variables $(v^r,p)$, it is necessary to reconstruct the primitive variables in terms of the conservative ones. During the evolution of the conservative variables, we recover the primitive variables exactly. Inspired in \cite{Choptuik}, we obtain that the primitive variables are in terms of the conservative ones

\begin{eqnarray}
\sqrt{\gamma}p&=&-2\sigma \tau +\sqrt{ 4\sigma^{2}\tau^{2} +  (\Gamma -1)(\tau^{2}-S^{2}) }, \\
v_{r}&=&\frac{S_{r}}{\tau+\sqrt{\gamma}p},
\end{eqnarray}

\noindent where $\sigma =\frac{2-\Gamma }{4}$ and $S^{2}=\gamma^{rr}S_{r}^{2}$. From this, once $p$ and $v_r$ are known, it is possible to reconstruct the energy density $\rho$ and the Lorentz factor $W$. 

{\it Boundary conditions.} In order to allow the fluid to enter the black hole we use Eddington-Finkelstein  horizon penetrating and choose the domain such that the black hole event horizon is contained in it, that is $r_{min}<r_{EH}$. At $r=r_{min}$ we apply the excision method \cite{SeidelSuen1992}, which can  be done since the surface $r=r_{min}$ is space-like, and in Eddington-Finkelstein type of coordinates the light cones all point toward the singularity and are open, then all the material arriving at such boundary will automatically get off the domain (or equivalently will fall toward the singularity) without the need of imposing boundary conditions there.  At the exterior boundary  $r=r_{max}$ we use radiative boundary conditions for the metric and extrinsic curvature components with background subtraction \cite{Thornburg}, while for the hydrodynamical variables, we use inflow boundary conditions. In order to avoid the contamination of the calculations, we locate the exterior boundary $r=r_{max}$ at a causally disconnected distance, such that our evolutions end before and potential noise coming from the boundary arrives at the black hole horizon.

% -> subsection
\subsection{Diagnostics}

{\it Apparent Horizon.} We are interested in tracking the growth of the black hole in time due to the accretion of the ultrarelativistic material. We thus track the apparent horizon, since it is a 2-surface (a two sphere in the spherically symmetric case) that can be located at each spatial hypersurface, that is, at every time step during the time integration of the equations. The apparent horizon is the outermost trapped surface satisfying

\begin{equation}
\Theta = \frac{\partial_r \gamma_{\theta\theta}}{\sqrt{\gamma_{rr}}\gamma_{\theta\theta}}-2\frac{K_{\theta\theta}}{\gamma_{\theta\theta}}=0, \label{eq:AH}
\end{equation}

\noindent where $\Theta$ is the expansion of the future pointing null vectors, whose projection is orthogonal and pointing outward the 2-spheres \cite{Baumgarte}. In order to track the apparent horizon, we calculate $\Theta$ at every time and locate the outermost zero of it at the coordinate radius $r_{AH}$. Then we can calculate the mass of the black hole apparent horizon $M_{AH}=R_{AH}/2$, where $R_{AH}=\sqrt{\gamma_{\theta \theta}(r_{AH})}$ is the areal radius evaluated at $r_{AH}$. We then track the growth in time of the apparent horizon radius which provides a good approximation of the mass of the black hole during the evolution.

{\it Constraints.} In order to validate the numerical solution of Einstein equations, the constraints (\ref{eq:constrains}) are required to be satisfied up to numerical errors. This is achieved by checking that the violation of the constraints converges to zero when the resolution of the numerical domain is increased. In order to show that the constraint is satisfied during the evolution one can calculate a norm of the violation at every time step. We calculate the $L_2$ norm of the constraint violation defined as $L_2(G) = \sqrt{\int |G|^2 d^3x}$, where the integral is performed numerically  in the spatial domain $r \in [r_{min},r_{max}]$. In our case we monitor $G=H$ and $G=M^r$.

{\it Units.} Both, Einstein and Euler equations are written assuming geometrical units $G=c=1$, which simplifies the calculation of the numerical solution. In this case, both $r$ and $t$ are in units of $M$. Knowing this, we set $M=1$ in the numerical solution. This units allow the calculation of the growth rate of the black hole horizon and estimate a final mass after a finite time.

When physical units are required at initial time when setting initial conditions, and at final time when calculating the final mass we proceed as follows. The radius of the initial black hole mass is twice its initial mass, which we define in solar masses. From there we use the radius in solar Schwarzschild radius $r_S{}_{\odot}$ in km. In this way, the spatial coordinate $r$ and the masses involved in the further analysis are thus in solar masses.

At initial time is is important to compare the spatial size of the black hole with  the cosmological particle horizon radius during the RDE, which is $r_{PH}=2 c t$. For that we calculate $t$, the cosmological time, in seconds.

% ----->     Initial data

\subsection{Initial Data}

In the ideal case we would solve the constraints (\ref{eq:constrains}) using an arbitrary gas distribution, for instance assuming a profile for the density as a source of the constraints. What is commonly done is to assume that the gas profile is localized in a bounded region, allowing the space-time to be asymptotically flat. We proceed in a different manner.

Since we plan to model a system that is not asymptotically flat consisting of a gas filling the entire space, and moving in a localized region, i) we start the evolution using (\ref{eq:gammaevolve}-\ref{eq:flux_conservative}) with an initially constant density profile of very low density, ii) constraints (\ref{eq:constrains}) are not satisfied initially, that is, they do not converge to zero initially, however the system gas plus space-time self-regulates and at a finite time the constraints converge from then on.

We parametrize the initial data with the initial value of the -initially- constant energy density profile $\rho_{ini}$. Such value of the density is also kept as the asymptotic value in our numerical domain, that we associate to the energy density of the cosmological environment. The other free parameter of the initial data is the radial velocity, which we parametrize with an asymptotic value $v_{\infty}$, normalized such that the radial velocity profile is $v^r=\frac{v_{\infty}}{\sqrt{\gamma_{rr}}}$. In all our simulations we have used the fixed value $v_{\infty}=0.9$.

% ----->     RESULTS     <-----

\section{Results on PBH mass growth}
\label{sec:results}

\subsection{Evolution of the black hole without considering the expansion of the universe}

As an example of one of our evolutions, we consider the case with $\rho_{ini}=10^{-9}$ in geometrical units and $M=1$. In Fig. \ref{fig:conv_horizon}. We show how quickly the system achieves a convergent regime. We also show the evolution of the apparent horizon mass.

For our analysis we then model the horizon mass growth as linear in time, that is, we fit the horizon mass in time with a fitting function $f=\dot{M}t +b$, where $\dot{M}$ is the growth rate of the black hole. Once we estimate $\dot{M}$ with sufficient accuracy, we use such parameter to estimate the energy density accreted during a given window of time $\Delta t$, that is $\Delta M=\dot{M} \Delta t$. The values of $\dot{M}$ for various values of the energy density appear in table \ref{tab:fits}.

\begin{figure}[htp]
\begin{center}
\includegraphics[width=7.5cm]{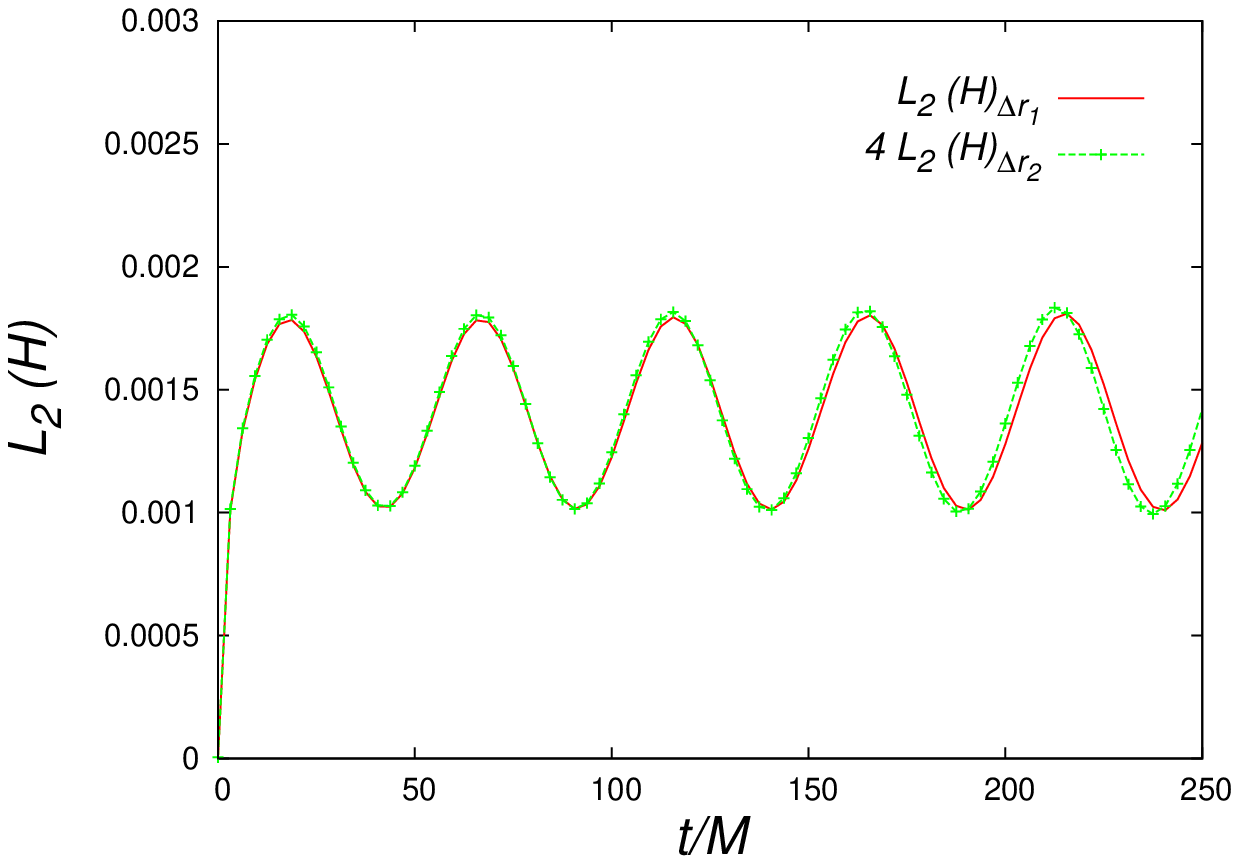}
\includegraphics[width=7.5cm]{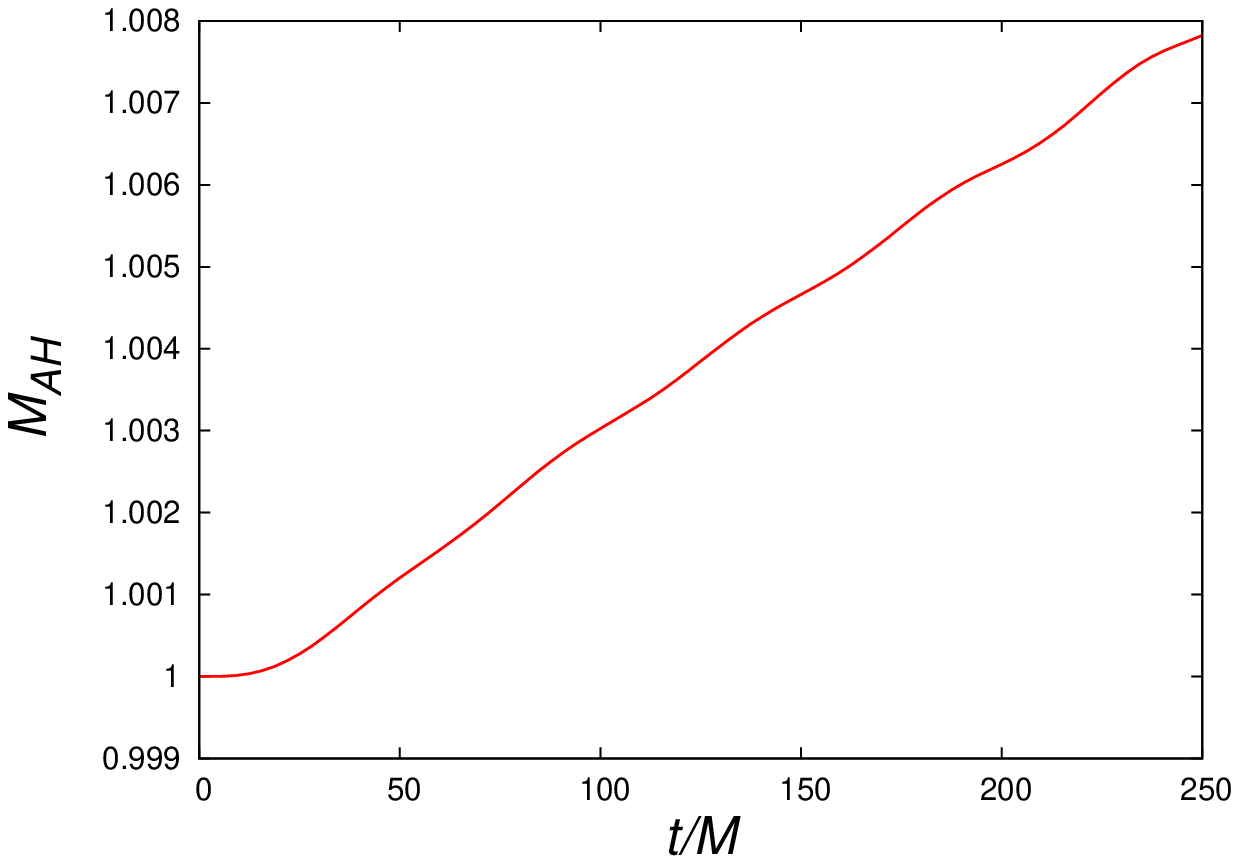}
\end{center}
\caption{\label{fig:conv_horizon} We show the second order convergence of the $L_2$ norm of the Hamiltonian constraint in time using two resolutions $\Delta r_1 = 0.0125M$ and $\Delta r_2 = \Delta r_1 /2$. The constraint calculated with the fine resolution has been scaled by the appropriate convergence factor $2^2$ and lies on top of the constraint calculated with the coarse resolution, which indicates second order convergence.  Even though the initial data are not consistent with the constraints at initial time, at about $t\sim 0.1M$ the system reaches a convergent regime. We also show the evolution of the Richardson extrapolation of the apparent horizon mass, that we calculated using the same two resolutions. The growth of the horizon is pretty linear in time after an initial transient.}
\end{figure}

We analyze the whole process of accretion as follows:

\begin{itemize}
\item[1)] Even though the initial density does not satisfy the constraints initially, they quickly become within the error and the convergence regime in about $\sim 0.1M$. This shows that our calculations are reliable almost immediately after we start the evolution.
\item[2)] After an initial transient, the black hole mass grows slowly, this is due to the fact that the density is redistributing and approaching a nearly stationary regime. After such initial transient the mass of the black hole starts growing in time linearly. This is the regime, between $t\in [50,250]M$ in the example of Fig. \ref{fig:conv_horizon}, where we guarantee numerical convergence and a pretty much stable behavior. There we fit the accreted mass with $\Delta M = \dot{M}t + b$, where $\dot{M}$ and $b$ are fitting parameters.
\item[3)] For the various densities in geometrical units used in our simulations, we find a linear relation between $\dot{M}$ and the density of the environment $\rho_{ini}$ shown in table \ref{tab:fits}. That is, we model this relation as $\rho_{ini} = c_1 \dot{M} + c_2$, where $c_1=3.193 \times 10^{-5}$ and $c_2=-5.96 \times 10^{-11}$ with errors smaller than $0.1\%$.
\end{itemize}

\begin{table}
\centering
 \begin{tabular}{ccc}\hline
  $\rho_{ini}$  	& $\dot{M}$ & $Error$ \\\hline\hline
  $10^{-9}$	& $3.32 \times 10^{-5}$ & $0.02\%$ \\ 
  $10^{-10}$	& $4.97 \times 10^{-6}$ & $0.04\%$ \\ 
  $10^{-11}$	& $2.19 \times 10^{-6}$ & $0.09\%$ \\ 
  $10^{-12}$	& $1.91 \times 10^{-6}$ & $0.10\%$ \\ 
  $10^{-13}$	& $1.88 \times 10^{-6}$ & $0.11\%$ \\ \hline
  \end{tabular}
  \caption{ \label{tab:fits} We show the fits of the black hole mass growth in time for various values of the asymptotic value of the energy density $\rho_{ini}$. The fits were carried out in the time interval $t \in [50M,250M]$.}
\end{table}

\subsection{Evolution of a sequence of stationary stages during the RDE era}

In order to consider the expansion of the universe in the evolution of the black hole we proceed as follows:

\begin{itemize}
\item[a)] We choose a time window within the RDE, specifically during the leptonic era, where we consider both, the equation of state is $p=\rho /3$ and the ultrarelativistic approximation holds. That is, we consider the universe mean energy density at the RDE goes like $\rho = K/t^2$; we fix the value of $K$ assuming that when the leptonic era starts at $t\sim 10^{-4}s$ the universe mean density is $\rho\sim 10^{16}kg/m^3$, then $K\sim 10^8 kg\cdot s^2/m^3$. In this way, we choose the energy density of the universe to be $\rho = 10^8 / t^2 ~[kg / m^3]$ during the leptonic time window $t\in [10^{-4} ~s,100 ~s]$. In this way, the value of $\rho$ introduces the contribution of the cosmic expansion.

\item[b)] We track the accretion process during this time domain as a sequence of stationary processes of accretion during finite time intervals $t\in[t_i,t_{i+1}]$ such that $t_0<t_1<...<t_{N-1}<t_f$. We choose the time intervals to be equally spaced in a logarithmic time scale, that is, we integrate the total accreted mass during the time $t\in[t_0,t_f]$ as the sum over the different time intervals.

\item[c)] Two illustrative examples of this process are shown in Fig. \ref{fig:superfigure}. For example, if the time interval $t\in[10^{-4}s,100s]$ is partitioned in six time intervals $t\in[10^{-4+i}s,10^{-4+i+1}s]$ for $i=0,1,...,6$. Thus, starting with a PBH mass $M^{PBH}_{0}=10^{-2}M_{\odot}$, the first interval is $t\in[10^{-4}s,10^{-3}s]$ using the density value $\rho=K/t^2 = 10^8/(10^{-4})^2kg/m^3 = 10^{16}kg/m^3$, then considering the accretion process is stationary as we discovered with our non-linear simulations, we calculate the accreted mass $\Delta M = 9.6 \times 10^{-3}M_{\odot}$, and then the black hole mass by the end of such interval is $M^{PBH}_{1}= 1.96 \times 10^{-2}M_{\odot}$. Then we consider a new stationary regime during the interval $t\in [10^{-3}s,10^{-2}s]$ with a density $\rho=K/t^2 = 10^8 / (10^{-3})^2 kg/m^3=10^{14}kg/m^3$, with the black hole mass $M^{PBH}_{1}= 1.96 \times 10^{-2}M_{\odot}$ and estimate $\Delta M$ and so on, until we cover the whole time domain up to $t=10^2s$. In Fig. \ref{fig:superfigure} we illustrate our algorithm for two cases of initial PBH masses of $M^{PBH}_{0}=10^{-2}M_{\odot}$ and $M^{PBH}_{0}=10^{-1}M_{\odot}$,  using only $N=6$.

\item[d)] We programmed a script that is able to iterate the process with a large number of time intervals $N$ in order to approach the continuum limit in time. We choose $N$ such that by increasing $N$ by two orders of magnitude the final mass of the black hole is the same to round-off error.
\end{itemize}

Based on these algorithms, we present the black hole growth using two time windows: case I) corresponding to $t\in[t^{-4}s,100s]$ in which it is assumed that the PBH was formed by the time $t\simeq10^{-4}s$ and case II) for $t\in[1,100s]$ in which we consider the PBH was formed at $t\simeq 1s$.

The results for Case I, including PBHs formed at $t=10^{-4}s$ are illustrated in Fig. \ref{fig:caseI}. The range of initial PBH mass is $M^{PBH}_0 \in [10^{-4},0.095]M_{\odot}$. The particle horizon at initial time is $r_{PH} =  2ct \sim 60km$ and the range of initial PBH masses covers the following range of black hole Schwarschild radius $r_{EH} \in [3\times 10^{-4},0.3]km$. For bigger values of $M^{PBH}_0$ we started to find non convergent results in terms of the number of time subintervals $N$. We associate this to the fact that the initial black hole radius approaches the particle horizon. Under the conditions of our analysis (especially the assumption of spherical flow), the threshold means that PBHs with masses bigger than this cannot be accurately calculated. The biggest PBH initial mass showing a finite final black hole mass is $M^{PBH}_{0} \sim 0.097M_{\odot}$. The main result of Case I, is that the masses of the final black hole lie on a scale of $50M_{\odot}$

On the other hand, for Case II, we find similar results, which are shown in Fig. \ref{fig:caseII}. The range of initial PBH mass is $M^{PBH}_0 \in [10^{-4},970]M_{\odot}$. The particle horizon at initial time is $r_{PH} =  2ct \sim 6\times 10^5 km$ and the range of initial PBH masses covers the following range of black hole Schwarschild radius $r_{EH} \in [3\times 10^{-4},3\times 10^3]km$. 
Again, a threshold for the initial PBH mass is found from which on we cannot obtain convergent results and associate to the fact that the Scharzschild radius of the initial black hole is approximately 1/100 of the particle horizon. For the range of masses showing a finite final black hole mass, we also find a lower limit of the final black hole mass of about $M^{PBH}_{f}\sim37M_{\odot}$ for all the initial black hole masses. The most massive black hole after $t\sim 100s$ is of the order of $10^6 M_{\odot}$ when the initial mass of the PBH is of the order of $900M_{\odot}$, which are already SMBHs. When the initial mass of the order of $M^{PBH}_{0}\sim1_{\odot}$, final black holes with masses of hundreds of solar masses are formed, which may well be SMBH seeds.

\begin{figure}[htp]
\includegraphics[width=8cm]{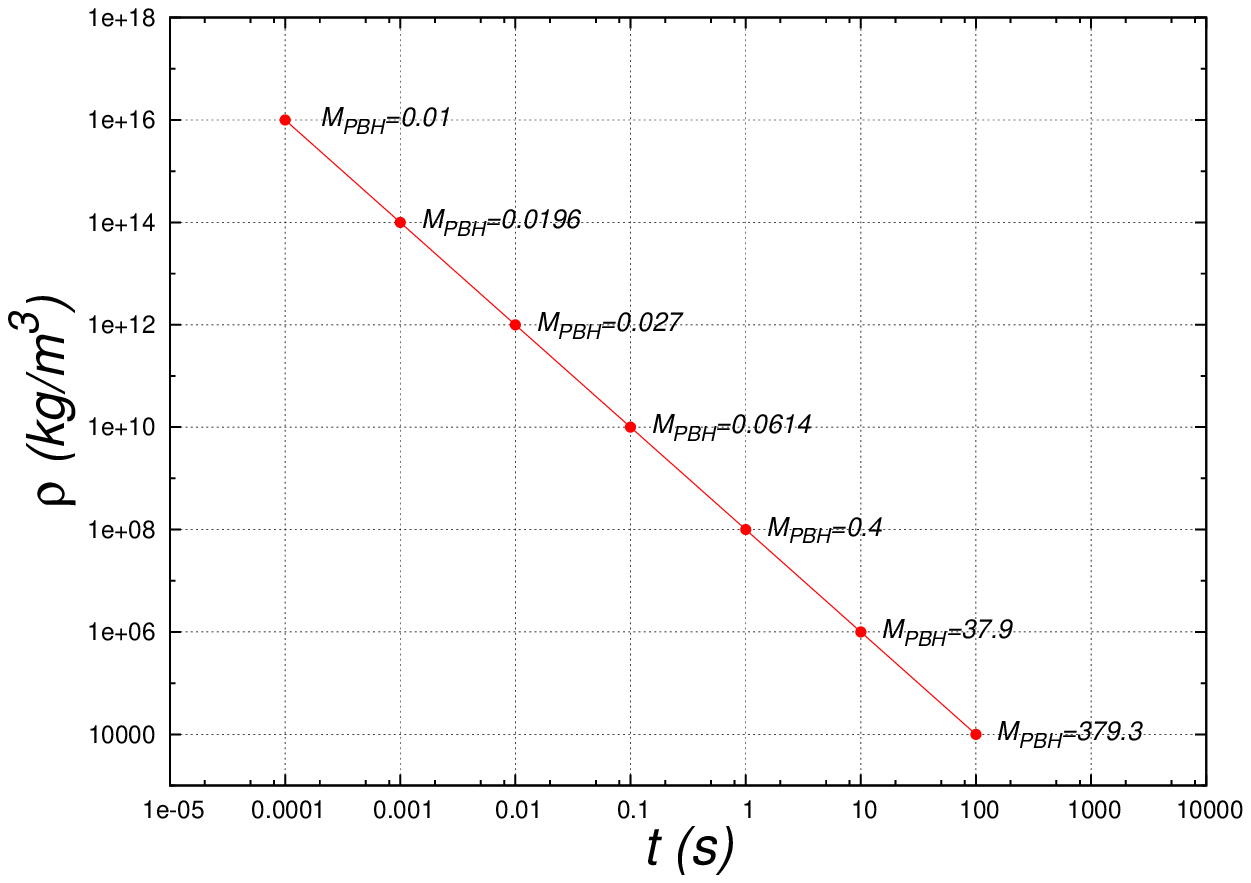}
\includegraphics[width=8cm]{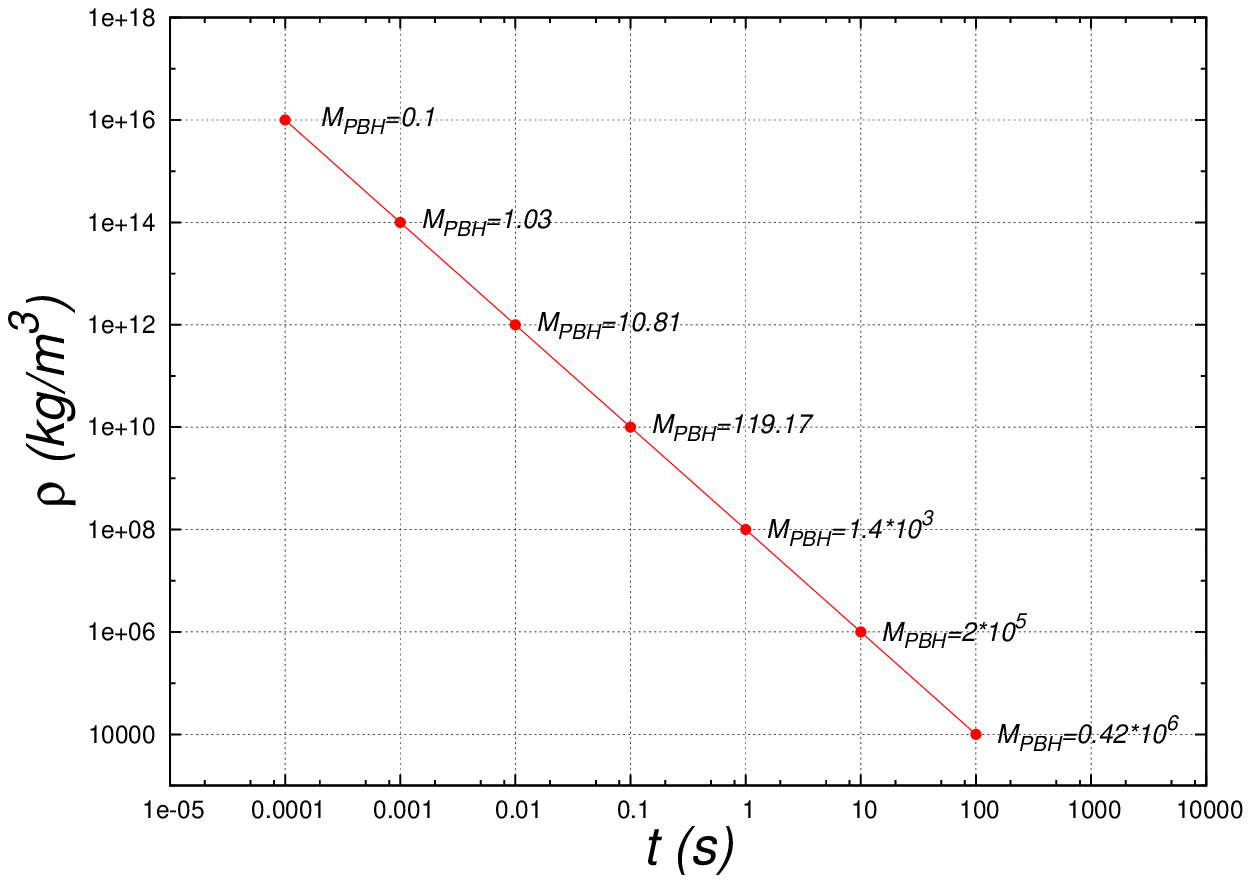}
\caption{\label{fig:superfigure} In this figure we illustrate our algorithm with two particular cases of a PBH growth in time using an extremely small number of time intervals $N=6$. The grid in the plot indicates also the time intervals we used to integrate the accreted mass $\Delta M = \dot{M} \Delta t$. In the first and second panels we show the evolution mass of a PBH with initial mass $M^{PBH}_{0}=0.01M_{\odot}$ and $M^{PBH}_{0}=0.1M_{\odot}$.}
\end{figure}

\begin{figure}[htp]
\includegraphics[width=14cm]{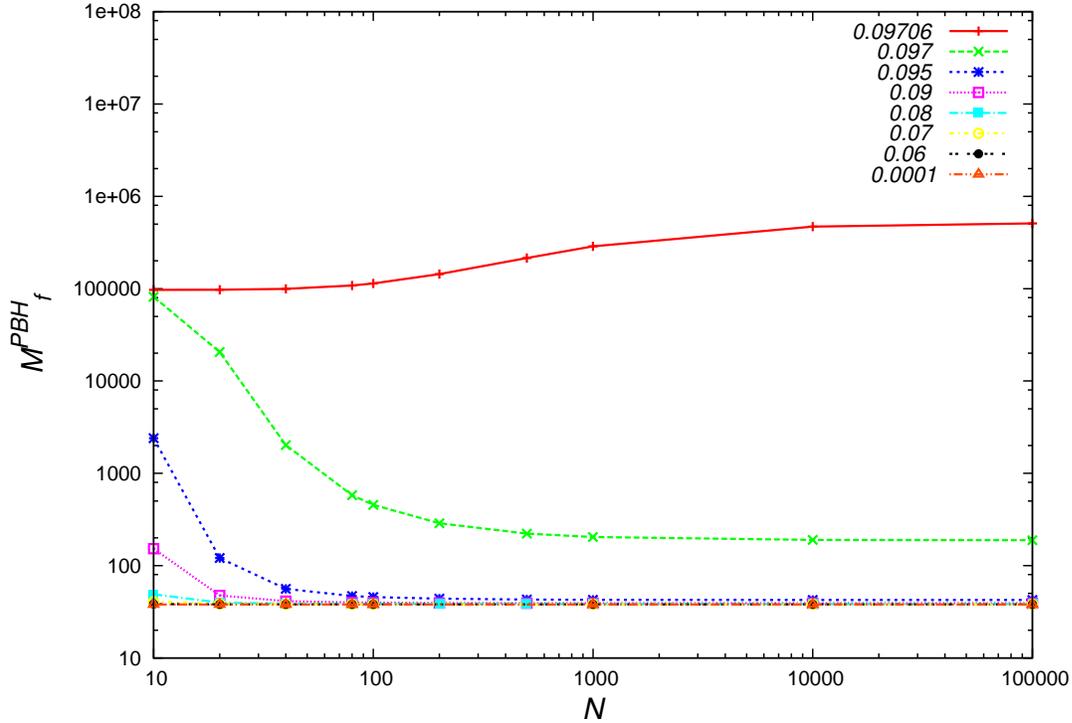}
\caption{\label{fig:caseI} We present the final mass of the PBH as a function of $N$ for various values of the initial mass of PBHs accreting during the time window $t\in [t^{-4}s,100s]$. First we show that the final mass stabilizes when increasing the number of time intervals $N$, which indicates that our calculation becomes independent of the time refinement level. Each line corresponds to the value of $M^{PBH}_{0}$ in solar masses. For initial PBH masses bigger than $0.09706M_{\odot}$ the mass calculations is not convergent anymore.}
\end{figure}

\begin{figure}[htp]
\includegraphics[width=14cm]{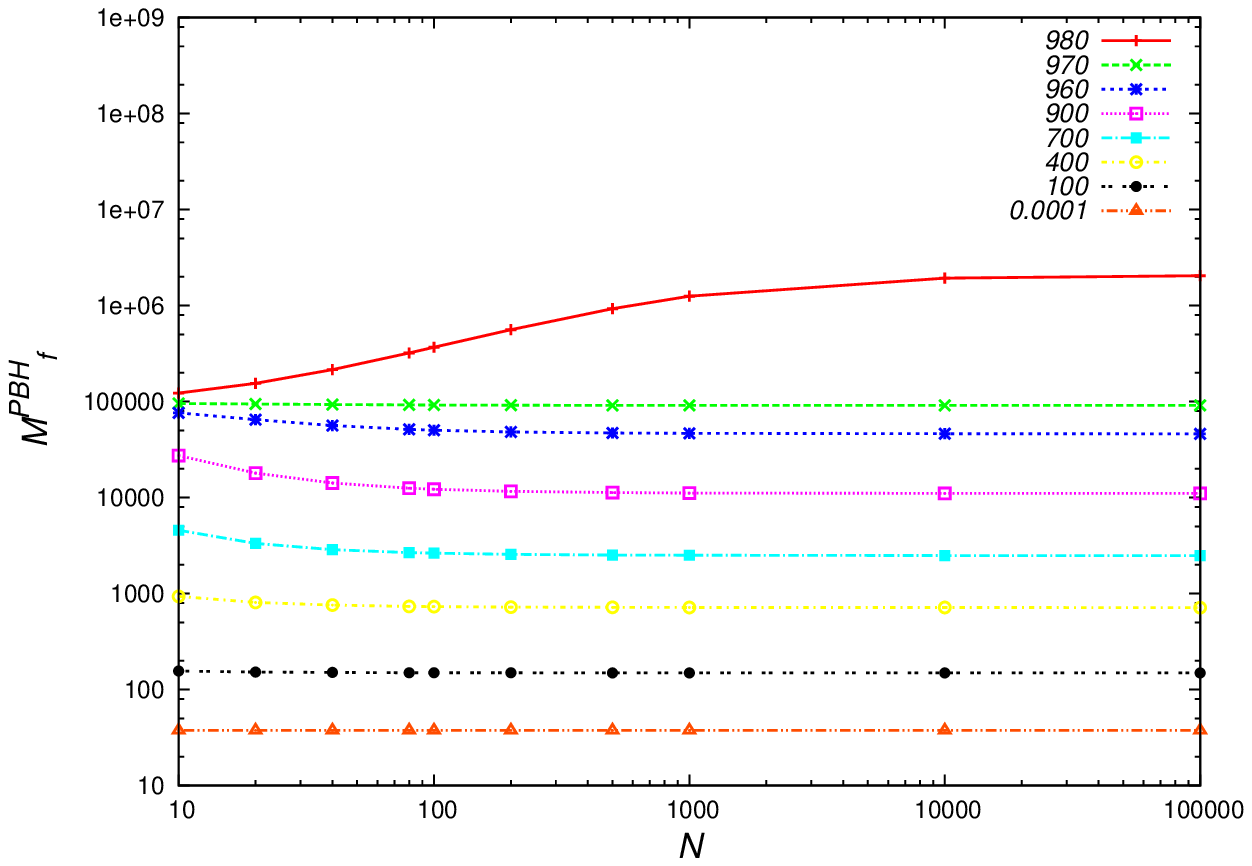}
\caption{\label{fig:caseII} We show the final mass for various values of the initial PBH mass $M^{PBH}_{0}$ in terms of $N$. We show the results for the Case II. Again, each line corresponds to the indicated value of the initial BH mass $M^{PBH}_{0}$ in solar masses. The cases shown correspond to finite values of $M^{PBH}_{f}$ and we show its convergence with $N$. For initial masses bigger that $980M_{\odot}$ the calculation is not convergent anymore.}
\end{figure}

% ----->     FINAL COMMENTS     <-----

\section{Discussion and Conclusions} 
 \label{sec:conclusions}

We modeled the radial accretion of ultrarelativistic gas using full non-linear numerical relativity and applied it to the growth of Primordial Black Holes, during the leptonic radiation dominated era, as a sequence of nearly stationary accretion stages. Various parameters are still free related to the formation and evolution of PBHs, one of them is the time window during which these object may accrete radiation. We use two time windows containing the leptonic era, where we consider our assumption of ultrarelativistic gas is valid. These two particular examples suffice to show that PBHs can grow up to seeds or SMBH masses during the RDE.

Analyzing a first time window, we found that if the PBH is formed at time $\sim 10^{-4}s$ PBH initial masses are required to be smaller than $M^{PBH}_{0} \sim 0.097M_{\odot}$, otherwise we do not find convergent results. When the PBH initial mass is smaller than this value, the mass of the black hole at time $100s$ is between $M^{PBH}_{f} \sim 35M_{\odot}$ and $ M^{PBH}_{f} \sim 200M_{\odot}$.

A second time window assuming the PBH was formed at time $\sim 1s$ shows that PBHs with initial masses between $M^{PBH}_{0} \sim 10^{-4}$ and $M^{PBH}_{0} \sim 0.1M_{\odot}$ accrete such that after 100$s$ they have masses bigger than $M^{PBH}_{f} \sim 37M_{\odot}$, whereas PBHs with initial masses between $M^{PBH}_{0} \sim 1M_{\odot}$ and $M^{PBH}_{0} \sim 100M_{\odot}$ accrete no more than a few hundreds of solar masses. A maximum initial PBH mass $M^{PBH}_{0} \sim980 M_{\odot}$ is allowed, after which we do not find convergent results. PBHs with initial masses between $900$ and $980M_{\odot}$ achieve masses between $10^4$ and $10^6M_{\odot}$, which may well be either SMBH seeds of SMBHs already formed respectively.

The fact that the most massive black holes formed at $t\sim 10^{-4}$s are small and thus acquire only a small mass is related to the fact that at the time of PBH formation, the Schwarzschild radius of the PBH is of the order of 1/100 of the particle horizon radius, whereas PBHs formed at $t\sim 1$s can have masses of the order of $\sim 900M_{\odot}$ and still have a radius 1/100 of the particle horizon. These later black holes are allowed to accrete about 1000 times their initial mass.

It is interesting to point out the contrast of our results with previous ones obtained in the past, specifically concluding that the accretion of radiation is not considerable \cite{Carr,HaradaCarrA}. The main new ingredient in our analysis is the incorporation of the solution of the Einstein-Euler system of equations,  during small time intervals. This is a considerable differences with previous models, for instance with the very first approach of Zel'dovich-Novikov \cite{Zeldovich} and the more modern ones in \cite{HaradaCarrA} where a wider class of equations of state are explored. Nevertheless, it remains the interesting question of why there is a threshold of non-convergent final black hole masses. For this we foresee an explanation. Consider the two limiting initial PBH masses for cases I and II described in the text,  in both cases $r_{PH}/r_{EH}\sim 200$. In the early Bondi accretion models on PBHs, there is such a threshold  of divergent final mass when the cosmic expansion is neglected, which for a radiation fluid occurs when initially $r_{PH}/r_{EH}=9\sqrt{3}/2\sim 7.8$  \cite{Zeldovich,HaradaCarrA,CarrReview}, or $r_{PH}/r_{EH}=3\sqrt{3}\sim 5.2$ according to \cite{Babichev,CarrReview}. Our threshold may perfectly be a general relativistic version of this threshold  when the initial ratio of particle horizon and black hole event horizon radii approaches $r_{PH}/r_{EH} \sim 200$.

Finally, even though our calculations involve the full non-linear solution of the Einstein-Euler system of equations, and a convergent sequence of successive stationary stages, we only consider the radial accretion. This is a limiting case of maximum accretion, showing the possibilities a PBH has to grow considerably and the bounds presented may change under different symmetry conditions of the flow and the black hole that are worth to investigate.

% ----->     ACKNOWLEDGMENTS     <-----

\section*{Acknowledgments}

We appreciate the comments form J. C. Hidalgo, and also the very constructive criticism from the anonymous referee. This research is partly supported by grants CIC-UMSNH-4.9 and CONACyT 106466.

% -------------------------------------------------------
% -----     REFERENCES     ----------
% -------------------------------------------------------

\end{document}